\documentclass[doublecol]{epl2}
\usepackage{epsfig}

\usepackage{amsmath}
\usepackage{graphicx}
\usepackage{color}

\newcommand{\be}{\begin{equation}}
\newcommand{\ee}{\end{equation}}
\newcommand{\bea}{\begin{eqnarray}}
\newcommand{\eea}{\end{eqnarray}}
\newcommand{\barr}{\begin{array}}
\newcommand{\earr}{\end{array}}

\newcommand{\ep}{\epsilon}
\newcommand{\al}{\alpha}

\newcommand{\bt}{\tilde{\beta}}

\title{Noise Rectification and Fluctuations of an Asymmetric Inelastic Piston }

\shorttitle{Noise Rectification and Fluctuations of an Asymmetric Inelastic Piston}

\date{\today}

\author{G. Costantini\inst{1} ,  U. Marini Bettolo Marconi\inst{1} \and A. Puglisi\inst{2}}

\institute{
\inst{1} Universit\`a di Camerino, Dipartimento di Fisica,
Via Madonna delle Carceri, I-62032 Camerino, Italy\\
\inst{2} Universit\`a di Roma ``La Sapienza'', CNISM and Dipartimento di Fisica, p.le Aldo Moro 2, I-00185 Roma, Italy
}

\pacs{05.40.-a}{Brownian motion}
\pacs{05.70.Ln}{Nonequilibrium and irreversible thermodynamics}
\pacs{45.70.-n}{Granular systems}

\abstract{We consider a massive inelastic piston, whose opposite faces have
different coefficients of restitution, moving under the action of
an infinitely dilute gas of hard disks maintained at a fixed temperature. 
The dynamics of the piston is Markovian and obeys a continuous Master
Equation: however, the asymmetry of restitution coefficients induces a
violation of detailed balance and a net drift of the piston, as in a
Brownian ratchet. Numerical investigations of such non-equilibrium
stationary state show that the velocity fluctuations of the piston are
symmetric around the mean value only in the limit of large piston
mass, while they are strongly asymmetric in the opposite limit. Only
taking into account such an asymmetry, i.e. including a
third parameter in addition to the mean and the variance of the
velocity distribution, it is possible to obtain 
a satisfactory analytical prediction for the ratchet drift velocity.}

\begin{document}

\maketitle

\section{Introduction}
Granular materials have been the subject of intense research in the
last 20 years in Physics~\cite{granulars}. Most of the non-trivial
phenomena that can be observed in a shaken box of sand are due to the
inelasticity of collisions among grains~\cite{gases}. Kinetic energy
is dissipated into heat, introducing an intrinsic time irreversibility
in the ``microscopic'' dynamics which can have consequences at a more
macroscopic level: for instance species segregation~\cite{unmix},
breakdown of energy equipartition~\cite{noneq}, apparent
Maxwell-demon-like properties such as heat currents against a
temperature gradient~\cite{nonmono} or mass current against a density
gradient~\cite{maxdem}, ratchet-like net drift of a asymmetrically
shaped tracers~\cite{noi}\cite{vdb1}, rectification of thermal fluctuations~\cite{vdb2}\cite{vdb3}, and so on. It is well known, moreover, that 
the space asymmetry of an adiabatic piston results in a stationary macroscopic 
motion \cite{Gruber}.
Here we consider, instead, a model of
asymmetric granular piston with different coefficients of restitution
which, with respect to a previously
presented model of granular Brownian ratchet~\cite{noi}, has the
advantages of being simpler to be studied analytically as well as
realized in the laboratory, and at the same time displays unexpected
peculiar properties: in particular we will show how its stationary
state is characterized by asymmetric velocity fluctuations, and that
this asymmetry becomes crucial when the piston mass is smaller than
the mass of surrounding disks. Such a light piston limit, which could
be regarded as purely academic in the case of a sect separating two
molecular gases, is instead realistic in the case of granular gases,
where the role ``molecules'' is played by large grains.

\section{Model}
The 2D model consists of a piston of mass $M$ and height $L$
surrounded by a dilute gas of $N$ hard disks of mass $m$ and density
$\rho$. The faces of the piston have two different values of
inelasticity, characterized by the coefficients of restitution
$\al_1$, on the left, and $\al_2$, on the right.  The piston can only
slide, without rotating, along the direction $x$, perpendicular to its
faces. The thickness of the piston can be neglected for the purpose of
our simulations and calculations.  The particles-piston binary
collisions are described by the rule:
\begin{eqnarray}
\label{eq:rule}
V &=& V'-(1+\al_{i})\frac{\ep^2}{1+\ep^2}(V'-v'_x)\nonumber \\
v_x &=& v'_x+(1+\al_{i})\frac{1}{1+\ep^2}(V'-v'_x)\nonumber \\
v_y &=& v'_y
\end{eqnarray}
where $\mathbf{v}$ and $\mathbf{v'}$ are the post-collisional and
pre-collisional disk velocities respectively, while $V$ and $V'$ are
the corresponding velocities of the piston, while $\ep^2=m/M$.
Because of the constraint on the piston, its vertical velocity is
always $0$. 
The energy of this granular system is not conserved and an external
driving mechanism is needed to attain a stationary state.  In our
model, the surrounding gas is coupled to a thermal bath which
keeps constant its temperature: the exact nature of the thermostat is
not discussed here (many models have been introduced in the
literature, see for example~\cite{kicks1,kicks2}), since we focus on the
dynamics of the piston which is assumed not to couple directly with
the thermostat, but only with the gas particles. For the sake of
simplicity we assume that the surrounding gas is homogeneous in space
and time, with a Maxwell-Boltzmann velocity probability density
function (pdf) $\phi(v_x,v_y)$ with zero average and given variance $\langle v_x^2 \rangle=\langle v_y^2 \rangle$
(see below). Such an assumption can be considered 
realistic as long as
the gas is dilute and the characteristic time of coupling with the
external heat bath is much shorter than that with the piston. In this
case also Molecular Chaos for piston-disks collisions can safely be
assumed, allowing for the use of the Direct Simulation Monte-Carlo
(DSMC) algorithm to simulate the piston dynamics~\cite{bird}.
Moreover, without loss of generality, we can choose one side of the
ratchet to be elastic. Based on the rules (\ref{eq:rule}), in fact, a
system with mass $M$ and coefficients of restitution
$0\leq\al_2<\al_1\leq 1$ is equivalent to one with $\al_1=1$ and
effective parameters $M'$ and $0\leq\al'_2\leq 1$ given by the
following relations
\begin{eqnarray}
M'&=&\frac{m(1-\al_1)+2M}{1+\al_1}\nonumber\\
\al'_2 &=& 1-2\frac{\al_1-\al_2}{1+\al_1}
\end{eqnarray}
In the following, therefore, we choose $\al_1=1$ and we 
study the behavior of the piston for different values of the mass
ratio $\ep^2$ and coefficient of restitution $\al_2$.

\section{Results and discussion}
Our choice of the initial conditions for the piston are $V(0)=0$ and
$X(0)=0$ After an initial transient whose duration depends on all
control parameters such as collision frequency, coefficients of
restitution, masses, etc., the piston reaches a stationary regime.  In
the following we shall indicate as $T_g=m \langle v_x^2 \rangle$ the
value of the gas temperature and $T_r=M\langle(V-\langle V
\rangle)^2\rangle$ the ratchet temperature.  The DSMC simulations have
been performed using $T_g=1$, $m=1$.
\begin{figure}[htbp]
\begin{center}
\includegraphics[angle=0,width=8cm,clip=true]{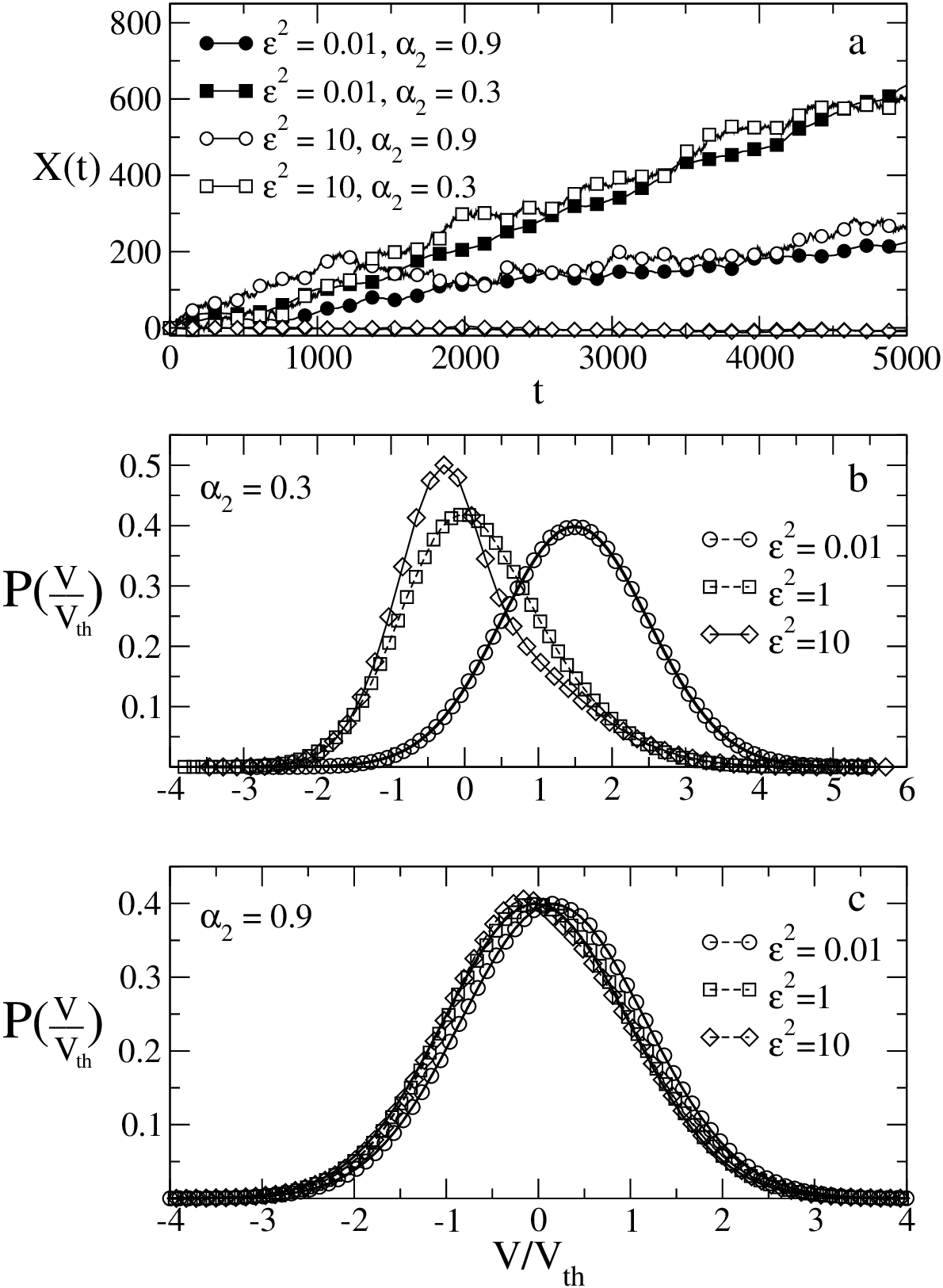}
\caption{Panel a: Single trajectories, in DSMC, of the massive piston
with $\ep^2=0.01$ (full symbol) and light piston with $\ep^2=10$ (open
symbol) for two different coefficients of restitution $\al_2=0.9$
(circles) and $\al_2=0.3$ (squares). Two cases with same
inelasticity ($\al_1=\al_2=0.3$) and different mass ($\ep^2=0.01$ and $\ep^2=10$)
are shown (diamonds). The
trajectories in these cases are averaged over $100$
realizations. Panel b and c: The rescaled probability density
functions, in DSMC, for the different mass of the piston: $\ep^2=0.01$
(circles) $\ep^2=1$ (squares) and $\ep^2=10$ (diamonds) for
$\al_2=0.3$ (panel b) and $\al_2=0.9$ (panel c). The velocity $V$ of
the ratchet is rescaled by thermal velocity $V_{th}=\sqrt{T_r/M}$. The
full line in the panel c represents a gaussian fit of the data.
\label{fig1}}
\end{center}
\end{figure}
We focus here on the behavior of the piston, whose position and
velocity at time $t$ are denoted as $X(t)$ and $V(t)$
respectively. Trajectories of the piston in its non-equilibrium
stationary state, i.e. discarding transients, for particular choices
of the parameters $M$ and $\al_2$, are displayed in Fig.~\ref{fig1}a.
When the system is totally elastic or $\al_1=\al_2$ no average motion
occurs. On the contrary, if the inelasticity of the left and right
side are different, the piston shows a mechanism of rectification of
disorder-induced fluctuations driven by the collisions and it drifts
with average velocity $\langle V \rangle \neq 0$. In particular this
drift appears to be always oriented towards the side of the piston
with smaller restitution coefficient, the positive direction in our
case, so that $\langle V \rangle >0$. 
This phenomenon can be understood by considering the average
momentum transferred by the gas to the piston, $M\langle V'-V \rangle$. 
Assuming that the piston is slower than the thermal velocity of
the gas it is easy to show that it will experience a viscous drag
force $-\gamma V$, where $\gamma$ is given by
$(2+\alpha_1+\alpha_2)\rho\sqrt{M/(M+m)}\sqrt{2 m T_g/\pi}$
plus a net force 
\be
F=\frac{\rho}{2} T_g \frac{M}{M+m} (\alpha_1-\alpha_2)
\ee
In other words, the momentum transferred by the gas
is smaller
on the more inelastic side, originating the drift in that
direction. 
From the simulation one also observes that the velocity of the drift
is bigger if the elasticity is smaller, while it seems not much
influenced by the piston mass. Mass, on the contrary, influences
fluctuations in the trajectories, which are larger the lighter is the
ratchet. In order to characterize the behavior of these fluctuations,
we show, in Fig.~\ref{fig1}b and Fig.~\ref{fig1}c, the pdf $P(V)$ of
the rescaled velocity of the piston.  When the piston is massive
$\ep^2\ll 1$, the distribution $P(V)$ does not display any appreciable
difference with respect to a Gaussian with finite mean. On the
contrary, when the mass $M$ of the piston is equal or smaller than
that of surrounding disks, $P(V)$ results asymmetric with a larger
tail for positive velocities. Such an effect occurs at smaller
elasticity and it is stronger if $\ep^2\gg 1$.  Note also that in all
our simulations we always have $\langle V \rangle \ge 0$, while in
some cases the maximum of the velocity pdf occurs at negative values.

In the dilute gas limit, it is possible to study the piston dynamics
by means of a Master Equation (ME) for $P(V,t)$ which can be written
as \cite{noi}: 
\be \label{eq:me} \frac{\partial P(V,t)}{\partial
t}=\int dV'~[W(V|V')P(V',t)-W(V'|V)P(V,t)] 
\ee 
where the transition
rate is: 
\begin{align} 
W(V|V')={}&\rho L \sum_{\substack {k=1}}^2 (-1)^k \int
d\mathbf{v} \Theta\big[(-1)^k(V'-v_x)\big]\nonumber \\
&(V'-v_x)\phi(\mathbf{v})
\delta\big[V-V'+\mu_k(V'-v_x)\big]
\label{boleq}
\end{align}
with $\mu_k=(1+\al_k)\ep^2/(1+\ep^2)$ (see Eq.~(\ref{eq:rule})) and
$\Theta$ is the Heaviside step function.

In order to obtain quantitative informations about the velocity and
granular temperature of the ratchet we can write a system of equations
describing the evolution of the first moments of the distribution,
starting from (\ref{eq:me}) and (\ref{boleq}), and invoke some
approximation to obtain a closed set.  By direct experience we have
found that, due to the asymmetry of $P(V)$, it is essential to
consider a further parameter $\xi=\langle (V-\langle V
\rangle)^3\rangle$ in addition to the average velocity and granular
temperature \cite{Sela}. To this aim we assume that the pdf of the
piston can be written as \bea P(V)=\sqrt{\frac{M}{2\pi
T_r}}\sum_{\substack{n=0}}^\infty a_n \frac{\partial^n}{\partial V^n}
\exp \Big[-\frac{M}{2T_r}(V-\langle V\rangle)^2\Big]
\label{pdfseries}
\eea
Imposing the normalization and that $\langle (V-\langle V\rangle)^2\rangle=T_r/M$, the series (\ref{pdfseries}), with a truncation at $n=3$, becomes
\bea
P(V)=\sqrt{\frac{M}{2\pi T_r}}\Big(1-\frac{\xi}{6}
\frac{\partial^3}{\partial V^3}\Big)
\exp \Big[-\frac{M}{2T_r}(V-\langle V\rangle)^2\Big]
\label{pdfnogauss}
\eea In the above expression $\xi$ represents the first term that is a
measure of the asymmetry of $P(V)$ about the average value. Under these assumptions the
equations for $\langle V \rangle$, $T_r$ and $\xi$ can be obtained
multiplying both sides of Eq.(\ref{eq:me}) by $V$, $M(V-\langle
V\rangle)^2$ and $(V-\langle V\rangle)^3$ respectively and performing
the integrations.  After some calculations, expressions in analytical
form for these equations can be derived if it is assumed that $\langle
V \rangle \ll V_{th}$, where $V_{th}=\sqrt{T_r/M}$. By retaining only
the terms of first order in $\langle V \rangle $, one obtains
the following differential equations 
\begin{align}
\label{avequations1}
\frac{\partial}{\partial t} \langle V \rangle=&{}
-\frac{\rho}{2}\frac{T_r}{M}a_1(\eta) - \rho\sqrt{\frac{2T_r}
{\pi M}} a_2(\eta)\langle V\rangle -\nonumber \\
&\frac{\rho}{3} \sqrt{\frac{M}
{2\pi T_r}} a_3(\eta) \xi \\
\label{avequations2}
\frac{\partial T_r}{\partial t} =&{}
\rho\sqrt{ \frac{2T_r^3}{\pi M} } b_1(\eta) + \frac{\rho}{2}
T_r b_2(\eta)\langle V\rangle +\frac{M\rho}{2} b_3 \xi \nonumber \\
&+\frac{\rho}{6} \sqrt{\frac{M^3}{2\pi T_r}} b_4(\eta) \langle V\rangle \xi
 \\
\label{avequations3}
\frac{\partial \xi}{\partial t} =&{}
-\rho \frac{3 T_r^2}{2 M^2}c_1(\eta)- \frac{\rho}{2} \sqrt{
\frac{2T_r^3}{\pi M^3} } c_2(\eta) \langle V \rangle +\nonumber \\
&\frac{\rho}{2}\sqrt{\frac{T_r}{2\pi M}} c_3(\eta) \xi + \frac{\rho}{2}
c_4\langle V\rangle \xi 
\end{align}
where $\eta=T_r/T_g$. The coefficients of the above equations are given in the appendix.\\

\begin{figure}[htbp]
\begin{center}
\includegraphics[angle=0,width=8cm,clip=true]{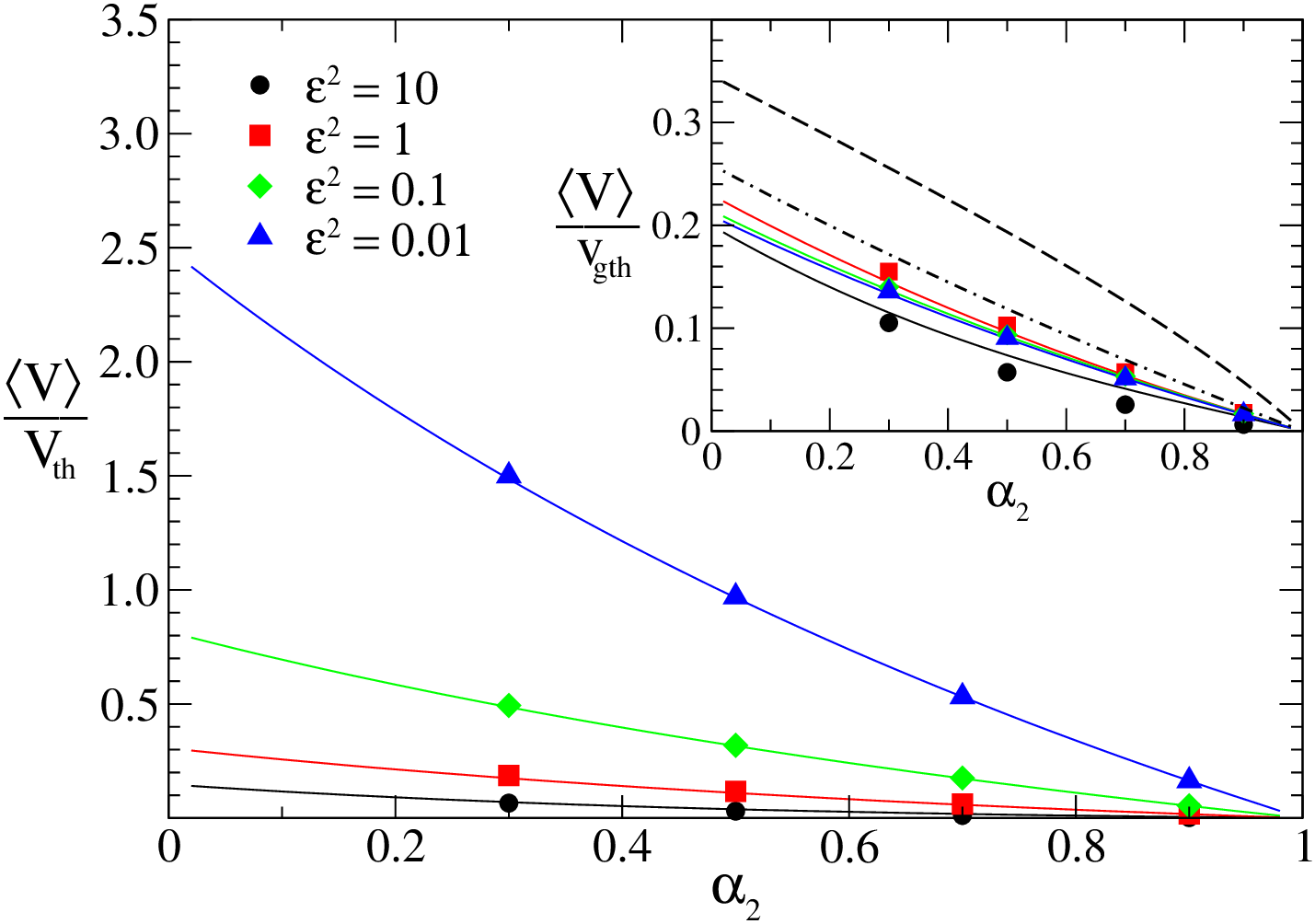}
\caption{The velocity of the piston, rescaled with its thermal velocity $V_{th}=\sqrt{T_r/M}$, as function of the coefficient of the restitution $\al_2$ for different values of the parameter $\ep^2$: $10$ (circles), $1$ (squares), $0.1$ (diamonds) and $0.01$ (triangles). The symbols correspond to the simulation data while the lines correspond to the solutions obtained from Eqs.(\ref{avev})-(\ref{eta}). Inset: The velocity of the piston, rescaled with the thermal velocity of the gas $v_{gth}=\sqrt{T_g/m}$, as function of the coefficient of the restitution $\al_2$ for the same values of $\ep^2$. The dashed and dot-dashed lines  correspond to the theoretical calculation for the cases $\ep^2=10$ and $\ep^2=1$ respectively, assuming $P(V)$ exactly gaussian.
\label{fig2}}
\end{center}
\end{figure}

\begin{figure}[htbp]
\begin{center}
\includegraphics[angle=0,width=8cm,clip=true]{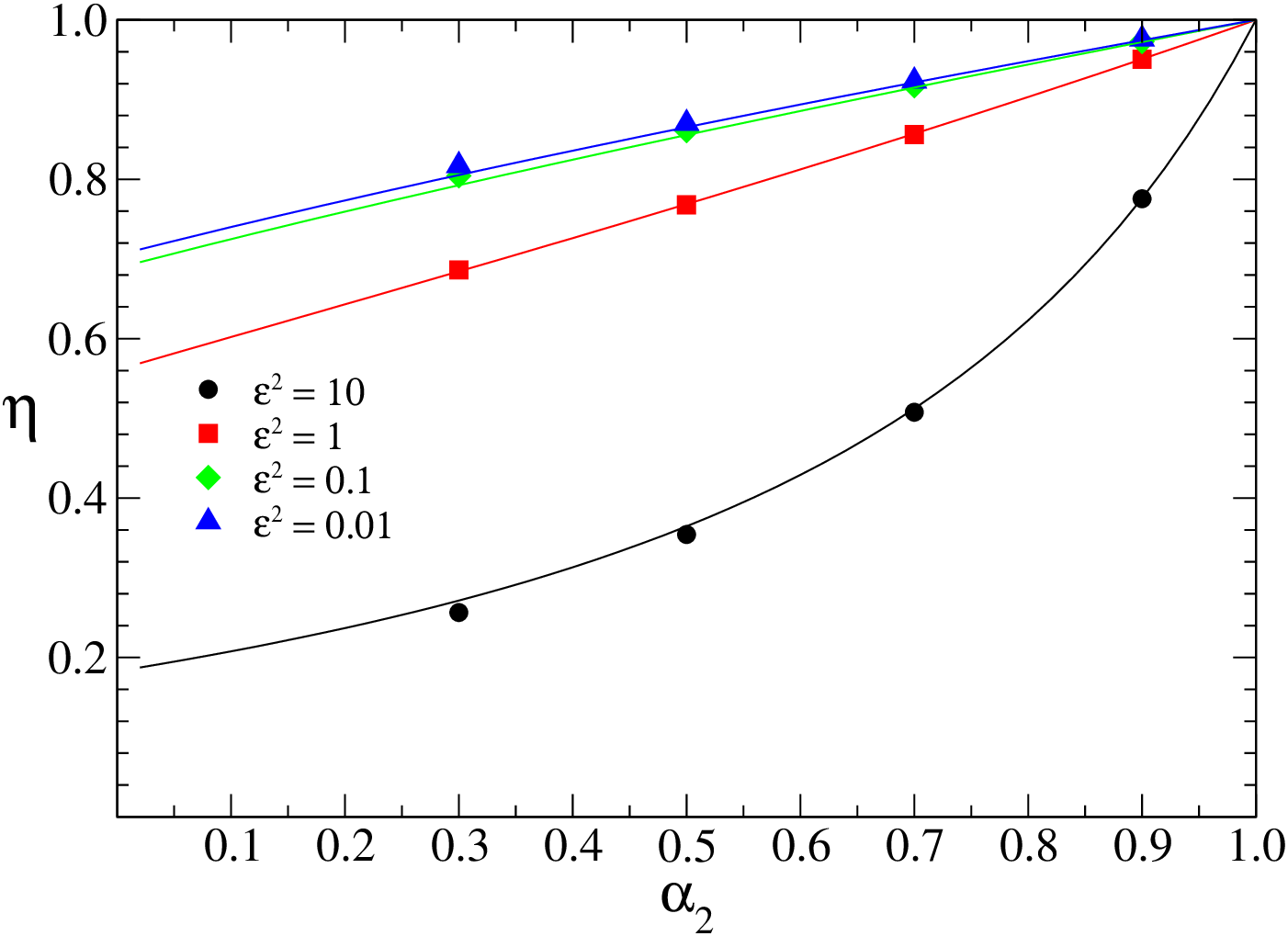}
\caption{The temperature ratio $\eta=T_r/T_g$ as function of the coefficient of the restitution $\al_2$ for the same cases of Fig. \ref{fig2}. The symbols correspond to the simulation data while the lines correspond to the solutions obtained from Eqs.(\ref{avev})-(\ref{eta}).
\label{fig3}}
\end{center}
\end{figure}

\begin{figure}[htbp]
\begin{center}
\includegraphics[angle=0,width=8cm,clip=true]{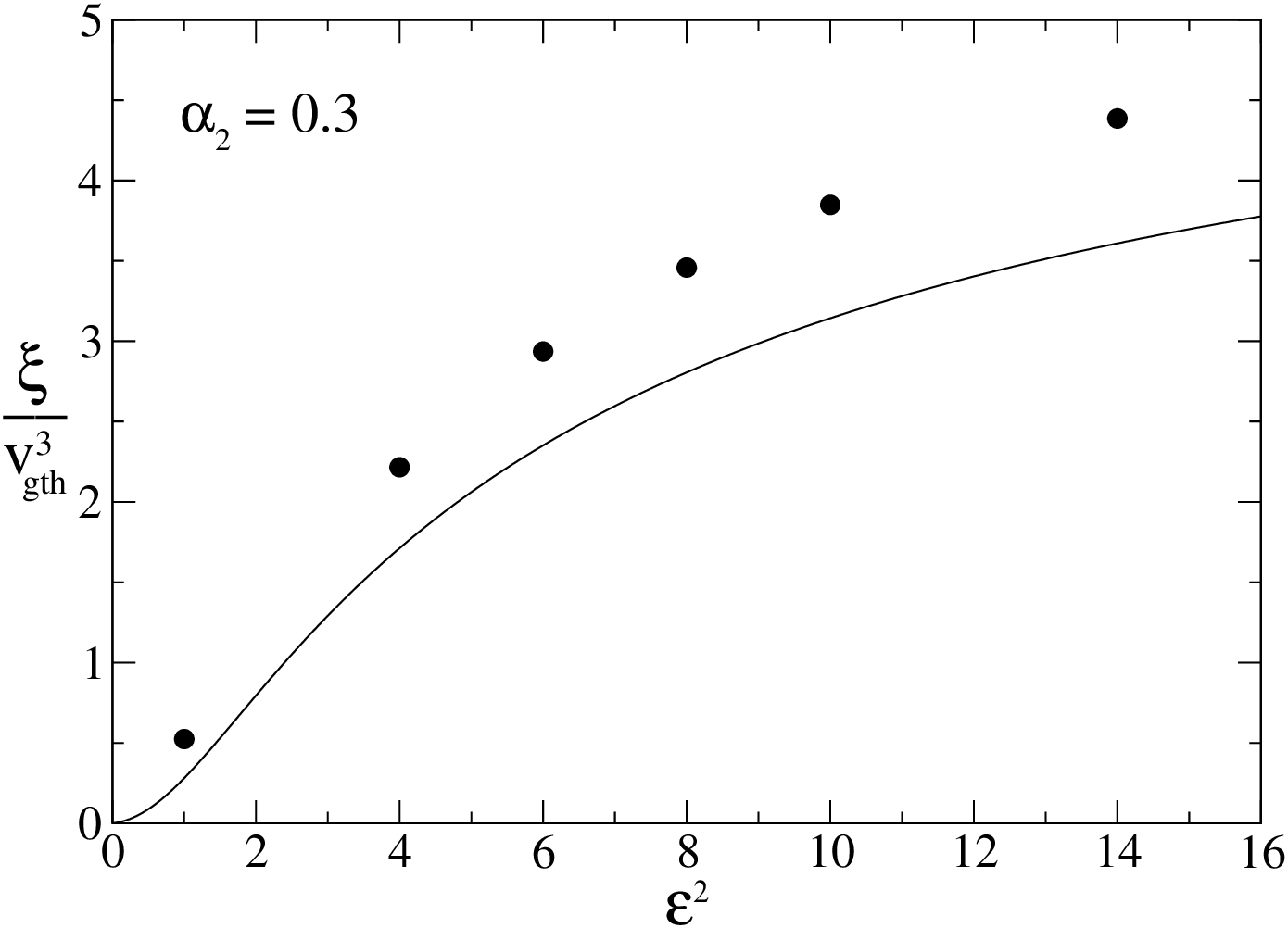}
\caption{The third moment $\xi=\langle (V- \langle V\rangle)^3\rangle$, rescaled with the quantity $v_{gth}^3$ as function of the parameter $\ep^2$ for $\al_2=0.3$. The symbols correspond to the simulation data while the line corresponds to the solution obtained from Eqs.(\ref{avev})-(\ref{eta})
\label{fig4}}
\end{center}
\end{figure}

We can solve Eqs.(\ref{avequations1})-(\ref{avequations3}) in the stationary state (all time derivatives are put to zero), obtaining:
\begin{align}
\label{avev}
&\langle V \rangle = -\frac{3}{2}\sqrt{\frac{\pi}{2}}V_{th}\frac{a_1(\eta)g_{34}(\eta) +2a_3(\eta)g_{14}(\eta)}
{3a_2(\eta)g_{34}(\eta)+ a_3(\eta)g_{24}(\eta)}\\
\label{csi}
&\xi = -\frac{3 V^2_{th}}{2 a_3(\eta)}[\sqrt{2\pi}V_{th}a_1(\eta)+4a_2(\eta) \langle V \rangle]\\
\label{eta}
&3\sqrt{2\pi}V_{th}[2 b_2(\eta)V^2_{th}\langle V \rangle +
2b_3\xi]+2b_4(\eta) \langle V \rangle \xi +\nonumber \\
&\quad 24 V^4_{th}b_1(\eta)=0 
\end{align}
where $g_{14}(\eta)=b_4(\eta) c_1(\eta) +4 b_1(\eta) c_4$, $g_{24}(\eta)=b_4(\eta) c_2(\eta) +3\pi b_2(\eta) c_4$ and $g_{34}(\eta)=b_4(\eta) c_3(\eta) -6\pi b_3 c_4$ (see appendix).\\
A numerical solution of system (\ref{avev})-(\ref{eta}) gives the stationary values of
$\langle V \rangle$, $\eta$ and $\xi$. In the
symmetric case $\al_1=\al_2=\al$, it turns out that
$a_1(\eta)=b_2(\eta)=b_3=c_1(\eta)=c_4=0$ the system has the solutions
$\langle V \rangle=\xi=0$ and from (\ref{eta}) we get
$\eta=(1+\al)/[\ep^2(1-\al)+2]$, that is the same solution already
obtained by Martin et al. \cite{Martin}. When $\al_1=\al_2=1$ the
temperature of the piston is equal to that of the gas.
The results in the asymmetric case are shown in Figs.(\ref{fig2})-(\ref{fig4}) together with the simulation data.
\begin{figure}[htbp]
\begin{center}
\includegraphics[angle=0,width=8cm,clip=true]{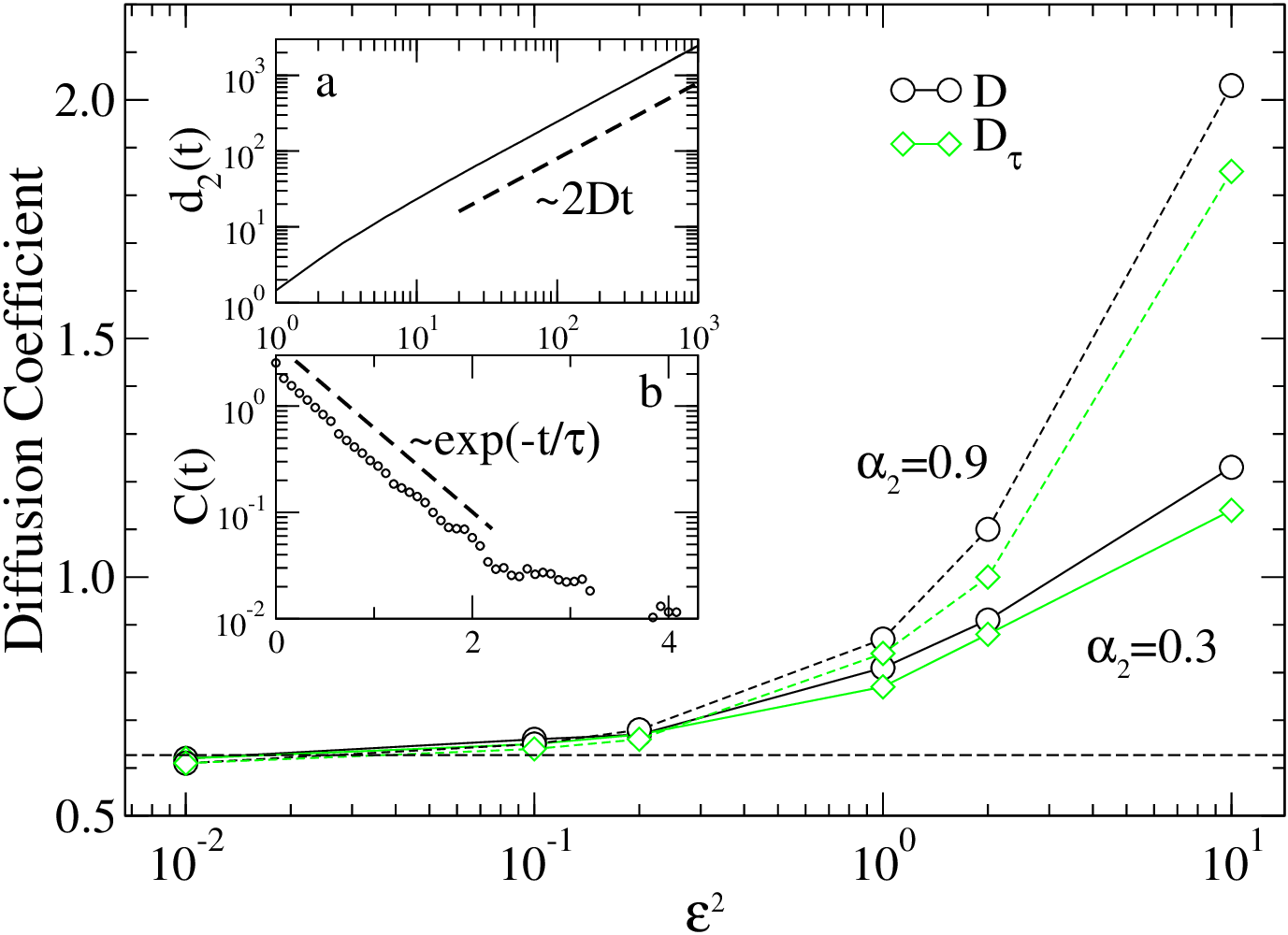}
\caption{The diffusion coefficient as a function of $\ep^2$ calculated
in two different ways (see text)for $\al_2=0.3$ and
$\al_2=0.9$. Inset: a) The self-diffusion $d_2(t)=\langle
(d(t)-\langle d(t) \rangle)^2 \rangle$ of the piston versus time; b)
The self-correlation function of the piston-velocity versus time. The dashed line corresponds to the theoretical prediction of $D_{\tau}$ (see text) for $\ep^2=0.01$, $\al_2=0.9$ and $\tau_c=2.5$.
\label{fig5}}
\end{center}
\end{figure}
Fig.~\ref{fig2} displays a good agreement between theory and DSMC for
the rescaled observable $\langle V\rangle/V_{th}$ for all values of
$\al_2$ and $\ep^2$ studied. It is remarkable that our results are
good even if the above assumption does not hold true, i.e. when
$\langle V \rangle \simeq V_{th}$. In the inset of Fig.~\ref{fig2} we
show the behavior of $\langle V\rangle/v_{gth}$ as a function of the
coefficient of restitution $\al_2$ for different $\ep^2$. The
theoretical results for ($M \le m$), derived assuming a Gaussian
$P(V)$ (i.e. $\xi=0$) are also displayed. They appear to disagree with
the numerical results, suggesting that in these cases a Gaussian pdf
for the piston velocity fluctuations is a very poor assumption. The
third moment $\xi$, that is a marker of the asymmetry of the pdf, is
not small (see Fig.\ref{fig4}) and expansion (\ref{pdfseries}) is
crucial. The small differences between theory and DSMC for $\ep^2=10$
(more visible in the inset of Fig. \ref{fig2}) are probably due to
the fact that we have considered only the first three terms of the
expansion. In fact the numerical results show for large values of
$\epsilon^2$ the fourth moment becomes relevant so that one should
take into account the evolution of such a quantity in the coupled set
of equations in order to give more accurate results.

We would like to add a comment concerning the high velocity tails of
the distribution function $P(V)$.  In spite of the fact that the body
of such a distribution is not Gaussian and skewed, we have been able
to prove that its tails are of Gaussian nature, but with different
effective temperatures in each tail, reflecting the asymmetry of the
piston. In particular, the temperature of each tail is given by a
formula similar to the one presented by Martin and Piasecki for the
inelastic intruder:
\be
T_i=T\frac{1+\alpha_i}{(1-\alpha_i) m/M+2}
\ee

Finally, we have studied, in Fig. \ref{fig5}, the diffusion coefficient $D$ of
the piston for different values of $\ep^2$. In order to better
characterize the situation of a light ratchet, we have calculated $D$
in two different ways. First we have measured it by its definition
$D=\lim_{t\to \infty}\frac{\langle (d(t)-\langle d(t) \rangle)^2
\rangle}{2t}$ where $d(t)=X(t)-X(0)$ is the displacement of the tracer
with respect to its initial position  at a  time $t=0$, taken when the whole system has
become stationary. 
Second we have used the self-correlation function
of the piston velocity defined by $C(t)=\langle (V(t)-\langle V
\rangle)(V(0)-\langle V \rangle) \rangle$: its time integral
$\int_0^\infty C(t') dt'$ gives an exact measure of the diffusion
coefficient. This time integral can also
be estimated by assuming an exponential decay $C(t)=T_r/M
\exp(-t/\tau)$: in this case $D_\tau=T_r
\tau/M$. 
The asymptotic decay, in fact, is always exponential, since the
process is Markovian by construction. This is checked in the insets of
Fig. \ref{fig5}. From Eqs. (\ref{avev})-(\ref{eta}) it can be deduced that, 
in the large $M/m$ limit the
dynamics of the piston is analogous to an Ornstein-Uhlenbeck process
with an effective friction $1/\tau$ and noise intensity $2T_r/\tau$,
with $\tau=\tau_c/[\ep^2(3+\al_2)]$ and $\tau_c$ is the average collision time. 
From such an expression it is
seen that the diffusion coefficient does not depend on the mass and very little on
the inelasticity. For smaller values of $M/m$, on the contrary, the
diffusion coefficient increases, as displayed in Fig. \ref{fig5}. 
A growing discrepancy between $D_\tau$ and $D$ is observed in this limit, 
due to the fact that the first part of the correlation decay is not exactly 
exponential.

\section{Conclusion}

We have discussed a simple example of noise rectification when two
identically driven granular gases are separated by a moving piston. 
The difference of piston inelasticites between the two faces, induces
a stationary drift and the appearance of non-Gaussian fluctuations,
which become highly asymmetric when inertia is reduced. A description
of the dynamics in terms of the first {\em three} moments of the
velocity distribution is sufficient to predict most of the physics of
this system. It is quite rare to reach such a detailed knowledge of a
system with asymmetric non-Gaussian fluctuations: this is achieved,
here, thanks to the empirical observation that the third cumulant of
the distribution is dominant over cumulants of higher order, a fact
that is not guaranteed in general~\cite{risken}. We are also confident
that the assumption used, mainly that the two gases on the sides of
the piston are very dilute and driven at high frequency, can be
reproduced in the laboratory and could be at the base of ratcheting
mechanisms observed in real granular materials.

\begin{widetext}
\section{A.1}
The coefficients of the Eqs. (\ref{avequations1})-(\ref{avequations3}) can be written as
\begin{align}
&a_1(\eta)=-\frac{2L(1-\bt)}{A^2(\eta)(1+\ep^2)}\\
&a_2(\eta)=\frac{2L\ep(1+\bt)}{A(\eta)(1+\ep^2)}\\
&a_3(\eta)=\frac{2L\ep^3(1+\bt)A(\eta)}{{1+\ep^2}}\\
&b_1(\eta)=\frac{4L\ep}{(1+\ep^2)^2} \Big[ 
\frac{1+\bt^2}{A^3(\eta)}-\frac{(1+\ep^2)(1+\bt)}{A(\eta)}\Big]\\
&b_2(\eta)=-\frac{4L\ep^2(1-\bt)}{(1+\ep^2)^2}\Big[\frac{3(1+\bt)}{A(\eta)}-2(1+\ep^2) \Big]\\ 
&b_3=-\frac{4L\ep^2(1-\bt)}{(1+\ep^2)^2}\Big(\ep^2\bt-1 \Big)\\
&b_4(\eta)=-\frac{8L\ep A^3(\eta)}{(1+\ep^2)^2} \Big\{2(1+\bt)\ep^2+ \ep^4(1-2\bt+3\bt^2)+\frac{3}{\eta} \big[\ep^2\bt^2-\bt(1+\ep^2)-1\Big]\Big\} \\
&c_1(\eta)=\frac{-2L(1-\bt)}{(1+\ep^2)^3}\Big\{3\ep^2-6\ep^4\bt+\ep^6(1-2\bt+4\bt^2)+
\frac{1}{\eta}\Big[1-2\ep^2(2+3\bt)+\ep^4(3+ 
2\bt+8\bt^2)\Big]+\frac{4\ep^2}{\eta^2}(1+\bt+\bt^2)\Big\} \\
&c_2(\eta)=\frac{4L\ep A^3(\eta)}{(1+\ep^2)^3} \Big\{ -2\ep^2+5\bt\ep^4-\ep^6(7-5\bt+8\bt^2)-\ep^8-\frac{3}{\eta}\Big[1-3\ep^2\bt+3\ep^4(3-2\bt+4\bt^2)-\ep^6(6+3\bt-12\bt^2+ \nonumber \\
&\quad 16\bt^3)\Big]+\frac{3}{\eta^2} \Big[1+\bt-2\ep^2(2-\bt+3\bt^2)+\ep^4(11+
\bt-6\bt^2+16\bt^3)\Big]+\frac{16\ep^2}{\eta^3}(1+\bt^3)\Big\}\\
&c_3(\eta)=\frac{-4L\ep A^3(\eta)}{(1+\ep^2)^3}\Big\{ 
4\ep^4\Big[3+3\bt+6\ep^2\bt(1-\bt)+\ep^4(1+3\bt-6\bt^2+4\bt^3)\Big]+\frac{\ep^2}{\eta}\Big[19(1+\bt)-2\ep^2(2-19\bt+21\bt^2)- \nonumber \\
&\quad \ep^4(29+19\bt+42\bt^2-32\bt^3)\Big]+\frac{1}{\eta^2} \Big[ 6(1+\bt)-6\ep^2(1-2\bt+ 
3\bt^2)+2\ep^4(2+3\bt-9\bt^2+8\bt^3)\Big] \Big\} \\
&c_4=\frac{4L\ep^2(1-\bt)}{(1+\ep^2)^3}\Big[3-3\ep^2(1+3\bt)+\ep^4(2-\bt+8\bt^2)\Big]
\label{coeff}
\end{align}
where $\bt=(1+\al_2)/2$ and $A(\eta)=\sqrt{\eta/(1+\eta\ep^2)}$.\\
\end{widetext}

\acknowledgments
 A. P. acknowledges COFIN-MIUR 2005 and  U.M.B.M. acknowledges a grant COFIN-MIUR 2005, 2005027808.


\end{document}